# EXACT SOLUTIONS OF THE KLEIN-GORDON EQUATION WITH HYLLERAAS POTENTIAL

Akpan N. Ikot[+1], Oladunjoye A. Awoga[1] and Benedict I. Ita [2]

[1] Theoretical Physics Group, Department of Physics, University of Uyo-Nigeria.

[2] Quantum Chemistry Group, Department of Chemistry University of Calabar-Nigeria

+ e-mail: ndemikot2005@yahoo.com

**Abstract**

We present the exact solution of the Klein-Gordon with Hylleraas Potential using the Nikiforov-Uvarov method. We obtain explicitly the bound state energy eigenvalues and the corresponding eigen function are also obtained and expressed in terms of Jacobi Polynomials.

1. **Introduction**

The study of the internal structure of any quantum mechanical systems such as nuclei, atom, molecules and solids form an integral part of different sets of orthogonal polynomials [1-5]. In nuclear and high energy physics the study of exact solution of the Klein-Gordon equation (KGE) is of high importance for mixed scalar and vector potentials. However, the problem of exact solution of the KGE for a number of special potential has been of great interests in the recent years. Different techniques have been assumed by different authors to obtain the exact or approximate solution of the KGE with some typical potential. These methods include the Standard Method [1] Supersymmetric Quantum Mechanics [2] Factorization Method [3], Asymptotic Iteration Method (AIM) [4], the Nikiforov-Uvarov (NU) Method [5] and others. The potentials investigated with these techniques include, Manning Rosen potential [6-7], Hulthen potential [8-9], Woods-Saxon potential [10-11], pseudoharmonic[12-13] ,Eckart potential [14-15]harmonic potential [16], Kratzer potential [17-18], Mie-type potential [19], ring – shape non-spherical oscillator [20], Hartmann potential [21], Scarf potential [22], Poschl-Teller potential [23], five parameters exponential – like potential [24] among others.

The relativistic Klein-Gordon equations contain two major objects: the four-vector linear momentum operator and the scalar rest mass. This allows one to introduce two types of potential, (i) the four vector potential $V(r)$ and (ii) the space time scalar potential $S(r)$. Alberto et al [25] have shown that when $S(r) = V(r)$ or $S(r) = -V(r)$, the KGE and Dirac equation share the same energy spectrum. The choice of $S(r) = V(r) = 2V(r)$ have been shown by Alhaidari et al [26-30] to yield a non relativistic limit for the KGE. However, analytical solutions of the KGE are only possible for the s-wave with the orbital angular momentum quantum number $l = 0$ for some potential [31-32]. On the other hand, the case $l \neq 0$, the KGE can only be solve approximately for some potentials using a suitable approximation techniques [33].

The Hylleraas potential is given by [34-35]

$$V(r) = D_e \left[1 - \frac{(1+a)(1+c)(s+b)}{(s+a)(s+c)(1+b)}\right] \quad (1)$$

where $D_e$ is the dissociation energy, with the function

$$s = e^{2(1+K)\omega r} \quad (2)$$

and the intermediates quantities $a, b, c$ are defined by

$$a = \left(\frac{K-k_2}{1+k_2}\right), \quad b = \left(\frac{K-k_1+k_2}{1+k_1+k_2}\right) \quad c = \frac{K-k_1}{1+k_1} \quad (3)$$

with $k_1, k_2$ and $K$ being the parameters. The Hylleraas potential is a special case of the multiparameter exponential type potential. This potential contains six parameters and it is very complicated and difficult to evaluate. Since Hylleraas [34] introduce this potential no much work have been reported on the bound state solution with this potential. Recently, we investigated the non-relativistic bound state solution with Hylleraas potential [36].

Motivated by the success in obtaining the bound state solution of the Schrödinger equation with the Hylleraas potentials. We are tempted to solve the solution of the KGE equation with the Hylleraas potential for $S(r) = V(r)$.

2. **Review of NU method**

The NU is based on solving the second order linear differential equation by reducing to a generalized equation of hyper-geometric type. This method has been used to solve the Schrödinger, Klein –Gordon and Dirac equation for different kind of potentials [28]. The second-order differential equation of the NU method has the form [5].

$$\psi''(r) + \frac{\bar{\tau}(s)}{\sigma(s)}\psi'(s) + \frac{\bar{\sigma}(s)}{\sigma^2(s)}\psi(s) = 0 \qquad (4)$$

where $\sigma(s)$ and $\bar{\sigma}(s)$, are polynomials at most second degree and $\bar{\tau}(s)$ is a first-degree polynomials. In order to find a particular solution to Eq. (4), we use the ansatz for the wave function as

$$\psi(s) = \varphi(s)\chi(s) \qquad (5)$$

It reduces Eq. (4) into an equation of hyper-geometric type

$$\sigma(s)\chi''(s) + \tau(s)\chi'(s) + \lambda\chi(s) = 0 \qquad (6)$$

and the other have function $\varphi(s)$ is defined as a logarithmic derivative

$$\frac{\varphi'(s)}{\varphi(s)} = \frac{\pi(s)}{\sigma(s)} \qquad (7)$$

The solution of the hyper-geometric type function in Eq. (6) are given by the Rodriques relation

$$\chi_n(s) = \frac{B_n}{\rho(s)}\frac{d^n}{ds^n}(\sigma^n(s)\rho(s)) \qquad (8)$$

where $B_n$ is a normalization constant and the weight function $\rho(s)$ must satisfy the condition

$$\frac{d}{ds}(\sigma\rho) = \tau\rho \qquad (9)$$

The function $\pi(s)$ and the parameter $\lambda$ required for the NU method are defined as follows:

$$\pi(s) = \frac{\sigma'-\bar{\tau}}{2} \pm \sqrt{\left(\frac{\sigma'-\bar{\tau}}{2}\right)^2 - \bar{\sigma} + k\sigma} \qquad (10)$$

$$\lambda = k + \pi' \qquad (11)$$

In order to find the value of $k$ in eq. (10), then the expression under the square root must be the square of the polynomials. Therefore, the new eigenvalue becomes

$$\lambda = \lambda_n = -n\tau' - \frac{n(n-1)}{2}\sigma'' \qquad (12)$$

where

$$\tau(s) = \bar{\tau} + 2\pi(s) \qquad (13)$$

whose derivative is negative. By comparing Eqs.(11) and (12), we obtain the energy eigen values

3. **Exact Bound State Solution of KGE**

In the relativistic atomic units ($\hbar = c = 1$) for a spindle particle, the three dimensional radial s – wave of KGE is [37]

$$[-\nabla^2 + (M + S(r))^2]\Psi(r,\theta,\varphi) = [E - V(r)]^2\Psi(r,\theta,\varphi) \qquad (14)$$

where $\nabla^2$ is the Laplacian and $E$ is the relativistic energy. The common ansatz for the wave function is $\Psi(r,\theta,\varphi) = R(r)/r \, Y_{lm}(\theta,\varphi)$, where $Y_{lm}(\theta,\varphi)$ is the spherical harmonic whose solutions is well-known [38]. Thus the radial part of the KGE becomes,

$$\frac{d^2 R(r)}{dr^2} + [(E^2 - M^2) + V^2(r) - S^2(r) - 2(EV(r) + MS(r))]R(r) = 0 \qquad (15)$$

Now considering the case of equal scalar and vector Hylleraas potential, Eq. (15) becomes

$$\frac{d^2 R(r)}{dr^2} + \left[(E^2 - M^2) - 2D_e(E + M)\left(1 - \frac{(1+a)(1+c)(s+b)}{(s+a)(s+c)(1+b)}\right)\right]R(r) = 0. \qquad (16)$$

Introducing the new variable, using Eq. (2), then it is straightforward to show that, Eq. (16) reduces to the form,

$$\frac{d^2 R}{ds^2} + \frac{2(1+c)(s+c)}{2(s+a)(s+c)(1+b)}\frac{dR}{ds} + \frac{1}{4(s+a)^2(s+c)^2(1+b)^2}[-\varepsilon^2 s^2 + \beta^2 s + \gamma^2]R(r) = 0 \qquad (17)$$

where we have used the following physical parameters:

$$\varepsilon^2 = \frac{-2\mu(1+b)\bar{E}}{(1+K)^2\omega^2} \qquad (18)$$

$$\beta^2 = \frac{2\mu(1+b)(a+c)E - (1+a)(1+c)\bar{V}}{(1+K)^2\omega^2} \qquad (19)$$

$$\gamma^2 = \frac{2(1+b)E - (1+a)(1+c)\bar{V}}{(1+K)^2\omega^2} \qquad (20)$$

where $\bar{E} = E^2 - M^2$ and $\bar{V} = 2D_e(E + M)$. Equation (17) is now amenable to NU method and in order to find the solution of this equation, it is necessary to compare Eq. (4) with Eq. (17). By this comparison, we obtain the following polynomials:

$$\bar{\tau}(s) = 2(1 + b)(s + c), \sigma(s) = 2(s + a)(s + c)(1 + b),$$

$$\bar{\sigma}(s) = -\varepsilon^2 s^2 + \beta^2 s + \gamma^2 \tag{21}$$

where

$$\beta^2 = \varepsilon^2 - \beta'^2 \tag{22}$$

$$\gamma^2 = \varepsilon^2 - \gamma'^2 \tag{23}$$

and $\beta'$ and $\gamma'$ are defined as

$$\beta'^2 = \frac{(1+a)(1+c)\bar{V}}{(1+K)^2\omega^2} \tag{24}$$

$$\gamma'^2 = \frac{(1+b(1+a)(1+c))\bar{V}}{(1+K)^2\omega^2} \tag{25}$$

respectively.

Substituting these polynomials into Eq. (10), we obtain the $\pi(s)$ function as

$$\pi(s) = \alpha_1(s + a) \pm \sqrt{(\alpha_1^2 + \varepsilon^2 + 2\alpha_1 k)s^2 + (\varepsilon^2 + \xi_1 + \alpha_1 k)s + \varepsilon^2 + \xi_2 + \alpha_3 k}, \tag{26}$$

where

$$\xi_1 = 2a\alpha_1^2 - \beta'^2 \tag{27}$$

$$\alpha_1 = (1 + b), \alpha_2 = 2\alpha_1(a + c) \tag{28}$$

$$\xi_2 = \alpha_1^2 a^2 - \gamma'^2, \alpha_3 = 2\alpha_1 ac \tag{29}$$

According to the NU method the expression in the square root must be square of the polynomial. Thus, we determine the complex $k - value$ as

$$= (\Lambda_2 + \Lambda_3 \varepsilon^2) \pm \sqrt{\left(\delta(\varepsilon^2 + A)\right)^2 - \left(\sqrt{A^2 - B}\right)^2}, \tag{30}$$

where

$$\Lambda_2 = 2\alpha_1 \xi_1 - 4\alpha_1^2 \alpha_3 - 8\alpha_1 \xi_2 \tag{31}$$

$$\Lambda_3 = (2\alpha_2 - 4\alpha_3 - 8\alpha_1) \tag{32}$$

$$A = \frac{2\Lambda_2\alpha_3 + 16\Lambda_1\alpha_1^2 + 16\Lambda_1\xi_1}{\delta^2} \tag{33}$$

$$B = \frac{\Lambda_2^2 - 4\Lambda_1\Lambda_4}{\delta^2} \tag{34}$$

$$\delta^2 = \Lambda_3^2 + 12\Lambda_1 \tag{35}$$

and $\Lambda_1$ and $\Lambda_4$ are defined as

$$\Lambda_1 = \alpha_2^2 - 8\alpha_1\alpha_3 \tag{36}$$

$$\Lambda_4 = \xi_1^2 - 4\alpha_1^2\xi_2 \tag{37}$$

respectively. The polynomial of $\pi(s)$ for each $k - value$ is

$$\pi(s) = \alpha_1(s + a) \pm \begin{cases} \sqrt{\left(\delta(\varepsilon^2 + A) + \sqrt{A^2 - B}\right)}s + \sqrt{\left(\delta(\varepsilon^2 + A) - \sqrt{A^2 - B}\right)} \\ for\ k = -(\Lambda_2 + \Lambda_3\varepsilon^2) + \sqrt{\left(\delta(\varepsilon^2 + A)\right)^2 - \left(\sqrt{A^2 - B}\right)^2} \\ \sqrt{\left(\delta(\varepsilon^2 + A) + \sqrt{A^2 - B}\right)}s - \sqrt{\left(\delta(\varepsilon^2 + A) - \sqrt{A^2 - B}\right)} \\ for\ k = -(\Lambda_2 + \Lambda_3\varepsilon^2) - \sqrt{\left(\delta(\varepsilon^2 + A)\right)^2 - \left(\sqrt{A^2 - B}\right)^2} \end{cases} \tag{38}$$

The correct value of $\pi(s)$ is chosen such that the function $\tau(s)$ has a negative derivative. Thus we select the $k$ and $\pi(s)$ values as

$$k = -(\Lambda_2 + \Lambda_3\varepsilon^2) - \sqrt{\left(\delta(\varepsilon^2 + A)\right)^2 - \left(\sqrt{A^2 - B}\right)^2} \tag{39}$$

$$\pi(s) = \alpha_1(s + a) - \left[\sqrt{\left(\delta(\varepsilon^2 + A) + \sqrt{A^2 - B}\right)}s + \sqrt{\left(\delta(\varepsilon^2 + A) + \sqrt{A^2 - B}\right)}\right] \tag{40}$$

which gives

$$\tau(s) = 2\alpha_1[2s + (a + c)] - 2\left[\sqrt{\left(\delta(\varepsilon^2 + A) + \sqrt{A^2 - B}\right)}s\right.$$

$$\left. - \sqrt{\delta(\varepsilon^2 + A) - \sqrt{A^2 - B}}\right] \tag{41}$$

and

$$\tau'(s) = -2\left[\sqrt{\delta(\varepsilon^2 + A) + \sqrt{A^2 - B}} - 4\alpha_1\right] < 0 \qquad . \qquad (42)$$

Using $\lambda = k + \pi'$ and its counterpart definition, $\lambda_n = -n\tau' - [(n(n-1)/2)]\sigma''$ given by Eqs. (12) and (13), the following expressions for $\lambda$ and $\lambda_n$ are obtained

$$\lambda = -(\Lambda_2 + \Lambda_3\varepsilon^2) - \sqrt{U^2 - V^2} \qquad (43)$$

$$\lambda_n = 2n\sqrt{U + V} - 2\alpha_1 n(n + 3) \qquad (44)$$

where $U^2 = \delta^2(\varepsilon^2 + A)^2$ and $V^2 = (A^2 - B)$.

Now using Eqs. (43) and (44), we obtain the complicated energy eigenvalues of the Klein-Gordon equation with Hylleraas potential as

$$\bar{E} = \frac{\omega^2(1+K)^2\sqrt{A^2-B}}{2\mu(1+b)\delta^2}\left[\Lambda_3 + \frac{A}{\sqrt{A^2-B}}\left(1 + \frac{\Lambda_3}{A}\sqrt{A^2-B} - \frac{\delta(1+2n)}{\sqrt{A^2-B}}\right)\right] \pm \frac{\omega^2(1+K)^2\sqrt{A^2-B}}{2\mu(1+b)\delta^2}$$

$$X \sqrt{\left(\left[\Lambda_3 + \frac{A}{\sqrt{A^2-B}}\left(1 + \frac{\Lambda_3}{A}\sqrt{A^2-B} - \frac{\delta}{\sqrt{A^2-B}}(1+2n)\right)\left(\frac{2\delta}{\sqrt{A^2-B}}\right)^2\left[\frac{A}{\sqrt{A^2-B}}\left(\frac{A}{2} - \frac{\delta(1+2n)}{\sqrt{A^2-B}}\right) + \alpha_1(1+2n(n+3) - \Lambda_3 - \sqrt{A^2-B}\right]^2\right)} \qquad (45)$$

Equation (45) is rather complicated transcendental equation. However, the energy eigenvalues can be found by setting the Hyllerass parameters in Appendix A into equation (45).

Let us turn attention to find the radial wave function for potential. Using $\tau(s), \pi(s)$ and $\sigma(s)$ in Eqs (7) and (9), we find

$$\varphi(s) = (s + a)^{\frac{F}{2}}(s + c)^{\frac{D}{2}} \qquad (46a)$$

where

$$D = \left(\frac{1-(\mu c+v)}{\alpha_1(c-a)}\right), F = \frac{\mu a+v}{2\alpha_1(c-a)} \qquad (46b)$$

and

$$\mu = \sqrt{\delta(\varepsilon^2 + A) + \sqrt{A^2 - B}}, \nu = \sqrt{\delta(\varepsilon^2 + A) - \sqrt{A^2 - B}}, \quad (46c)$$

$$\rho(s) = (s + a)^F (s + c)^D \quad (47)$$

Then from Eq. (8), we obtain

$$\chi_n(s) = B_n (s + a)^{-D} (s + c)^{-F} \frac{d^n}{ds^n} (s + a)^{n+D} (s + c)^{n+F} \quad (48)$$

and the total wave function can be written in the form

$$\psi(r, \theta, \varphi) = N_n \frac{1}{r} [a + exp(2(1 + K)\omega r)]^{\frac{D}{2}} [c + exp(2(1 + K)\omega r)]^{\frac{F}{2}} P_n^{(D,F)}(r) Y_{lm}(\theta, \varphi) \quad (49)$$

where $N_n$ is normalization constant and $P_n^{(D,F)}(r)$ is the Jacobi polynomials. The wave function obeys the normalization condition, $\int_{-\infty}^{\infty} r^2 |\psi(r, \theta, \varphi)|^2 \, d\tau$, where $d\tau$ is the volume element.

## 6.    Conclusion

In this work, we used the NU method for Hylleraas potential to obtain the exact solution of the Klein-Gordon equation. We obtain the exact energy eigen values equation and the unnormalized radial wave function expressed in terms of the hypergeometric function.

Finally, this result will open ways for more research work on the bound state solution of the Klein-Gordon equation with the six parameter Hylleraas potential and this will leads many authors to study the bound state energy spectra of some diatomic molecules with Hylleraas potential [34-35].

## Appendix A: Values of formulas in terms of the Hylleraas potential

The following are the useful formulas and relation satisfied by the energy eigen values of the Hylleraas potential of Eq. (45). The $\alpha_1, \alpha_2, \alpha_3$ are given in terms of Hylleraas parameters as

$$\alpha_1 = (1 + b) \quad (A.1)$$

$$\alpha_2 = 2(1 + b)(a + c) \quad (A.2)$$

$$\alpha_3 = 2ac(1 + b) \quad (A.3)$$

The expression $\beta'^2$ is defined as

$$\beta'^2 = \frac{(1+a)(1+c)\bar{v}}{(1+K)^2\omega^2} \tag{A.4}$$

The parameters $\Lambda_1, \Lambda_2, \Lambda_3, \Lambda_4$ are

$$\Lambda_1 = 4(1+b)^2(a^2 - 14ac + c^2) \tag{A.5}$$

$$\Lambda_2 = 4a(1+b)^3(2a - c + 1)$$
$$- \frac{2(1+a)(1+b)(1+c)(4b-3)\bar{v}}{(1+K)^2\omega^2} \tag{A.6}$$

$$\Lambda_3 = 4(1+b)(a + c - 2ac - 2) \tag{A.7}$$

$$\Lambda_4 = \frac{(1+a)(1+c)\bar{v}}{(1+K)^2\omega^2}\left[b(1+b)^2 + \frac{(1+a)(1+c)\bar{v}}{(1+K)^2\omega^2}\right] \tag{A.8}$$

Other essentials parameters expressed in terms of the Hylleraas parameters are defined as follows:

$$\delta = 64(1+b)^2[a(1-c) - a(8c+1) + c^2(1-a) + a + 4] \tag{A.9}$$

$$\xi_1 = 2a(1+b)^2 - \frac{(1+a)(1+c)\bar{v}}{(1+K)^2\omega^2} \tag{A.10}$$

$$\xi_1 = a^2(1+b)^2 - \frac{(1+b)(1+a)(1+c)\bar{v}}{(1+K)^2\omega^2} \tag{A.11}$$

$$\gamma'^2 = \frac{(1+b)(1+a)(1+c)\bar{v}}{(1+K)^2\omega^2} \tag{A.12}$$

Finally, the constants $A$ and $B$ are defined as

$$A = \frac{2(1+b)^2(16a^3+12a^2c-a^2c^263a^2c+8a^2-12ac+16ac^2+8c^2)\frac{(1+a)(1+c)}{(1+K)^2\omega^2}(16a^2+4bc-ac^2-64ac+16c^2)\bar{v}}{1024(1+b)^2[a^2-a^2c-8ac-a+4-ac^2-c+c^2]^2}$$

(A.13)

$$B = \frac{a(1+b)^4(2a-c+1)-\frac{2(1+b)^2(1+c)\bar{v}}{(1+K)^2\omega^2}(2a^2b+2bc^2-28abc+4ab-3a)+\left[\frac{(1+a)(1+c)\bar{v}}{(1+K)^2\omega^2}\right]^2(16b^2-4a^24c^2-56ac+a-24b)}{1024(1+b)^2[a^2-a^2c-8ac-a+4-ac^2-c+c^2]^2}$$

(A.14)